\documentclass[12pt,letterpaper]{article}

\usepackage{lipsum}
\usepackage[left=1 in,top=1in, right=1 in, bottom=1.1in]{geometry}

\usepackage{graphicx}
\usepackage{booktabs}
\usepackage{multirow}
\usepackage{setspace}
\usepackage{url}
\usepackage[most]{tcolorbox}
\newtcolorbox{remarkbox}{
  colback=white, colframe=black, 
  boxrule=0.5pt, arc=2pt, 
  left=6pt, right=6pt, top=3pt, bottom=3pt,
  title=Remark
}
\usepackage{crimson}
\usepackage[T1]{fontenc}
\usepackage{color}
\definecolor{darkred}{RGB}{117,0,20}

\definecolor{mybrown}{RGB}{139, 69, 19}

\usepackage{titlesec}
\titleformat{\section}
  {\normalfont\fontsize{16}{0}\bfseries}{\thesection}{1em}{}
\titleformat{\subsection}
  {\normalfont\fontsize{14}{0}\bfseries}{\thesubsection}{1em}{}  % was \thesection
\titleformat{\subsubsection}
  {\normalfont\fontsize{12}{0}\bfseries}{\thesubsubsection}{1em}{}  % was \thesection
\titleformat{\paragraph}
  {\normalfont\fontsize{12}{0}\bfseries\itshape}{\theparagraph}{1em}{}
\titleformat{\subparagraph}
  {\normalfont\fontsize{12}{0}\itshape}{\thesubparagraph}{1em}{}  % was \theparagraph
\makeatletter
\newcounter{subsubparagraph}[subparagraph]
\renewcommand\thesubsubparagraph{%
  \thesubparagraph.\@arabic\c@subsubparagraph}
\newcommand\subsubparagraph{%
  \@startsection{subsubparagraph}    % counter
    {6}                              % level
    {\parindent}                     % indent
    {12pt} % beforeskip
    {6pt}                           % afterskip
    {\normalfont\fontsize{12}{0}}}
\newcommand\l@subsubparagraph{\@dottedtocline{6}{10em}{5em}}
\newcommand{\subsubparagraphmark}[1]{}
\makeatother
\titlespacing*{\section}{0pt}{10pt}{4pt}
\titlespacing*{\subsection}{0pt}{6pt}{0pt}
\titlespacing*{\subsubsection}{0pt}{6pt}{1pt}
\titlespacing*{\paragraph}{0pt}{3pt}{3pt}
\titlespacing*{\subparagraph}{0pt}{3pt}{3pt}
\titlespacing*{\subsubparagraph}{0pt}{3pt}{3pt}

\usepackage[tableposition=top, figureposition=bottom, font=normalsize, labelfont=bf]{caption}

\usepackage{fancyhdr}
\usepackage[scaled]{helvet}

\makeatletter
\renewcommand{\maketitle}{\bgroup
   \begin{center}
   \textbf{{\fontsize{19.8pt}{36}\selectfont \textsf{\@title}}}\\
   \vspace{10pt}
   {\fontsize{12pt}{0}\selectfont \@author} 
   \end{center}
}
\makeatother

\usepackage{float}

\usepackage{enumitem}
\usepackage{graphicx}
\usepackage{eso-pic}  % for background elements
\fancypagestyle{firstpage}{
\fancyhf{}
\fancyhead[L]{%
  \includegraphics[height=0.7cm]{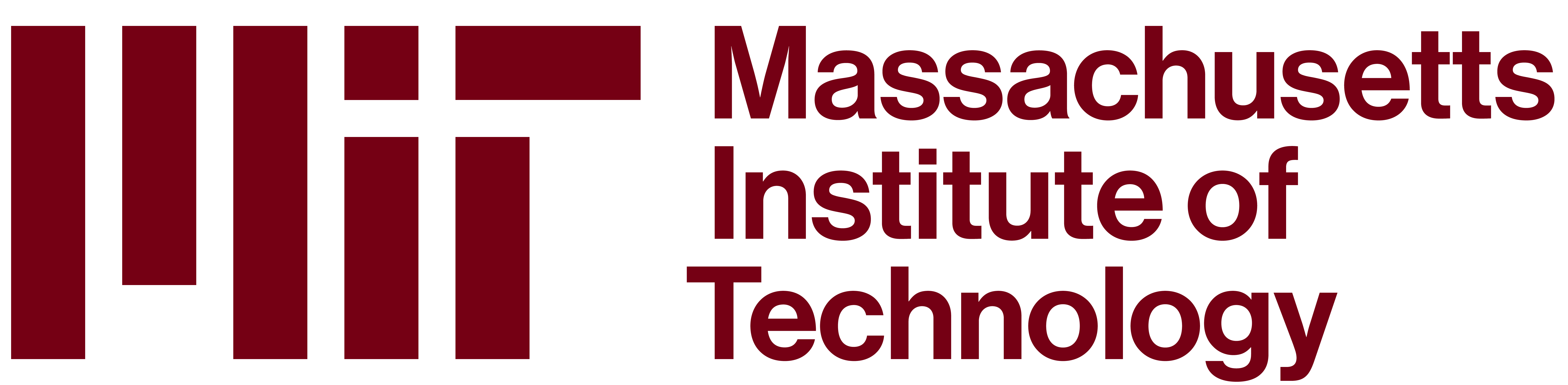}
}
\fancyhead[R]{%
  \textbf{OSGym}
}
\fancyfoot[C]{\thepage}
}

\fancypagestyle{fancy}{
  \fancyhf{}
  \fancyhead[C]{\footnotesize OSGym: Scalable OS Infra for Computer Use Agents}

  \fancyfoot[C]{\thepage}
}

\newenvironment{myquote}[1]%
  {\list{}{\leftmargin=#1\rightmargin=#1}\item[]}%
  {\endlist}
  
\renewenvironment{abstract}
{\vspace*{-.5in}\fontsize{12pt}{12}\begin{myquote}{0.3in}
\noindent \par{\bfseries \abstractname.}}
{\medskip\noindent
\end{myquote}
}
\begin{document}

% Set Title, Author, and email
\title{OSGym: Scalable \textcolor{darkred}{OS Infra} for \textcolor{darkred}{Computer Use} Agents}

\author{%
  Zengyi Qin$^{1\alpha\nu}$ ~
  Jinyuan Chen$^{\alpha}$ ~
  Yunze Man$^{2\alpha}$ ~
  Shengcao Cao$^{2\alpha}$ ~
  Ziqi Pang$^{2\alpha}$ ~
  Zhuoyuan Wang$^{3\alpha}$ \\ [0.3em]
  Han Fang ~
  Ling Zhu ~
  Zixin Xie ~
  Zibu Wei ~
  Tianshu Ran ~
  Haoran Geng$^{6}$ ~
  Ray Pan ~
  Qizhen Sun \\ [0.3em]
  Zachary Bright ~
  Yuyang Cai ~
  Chongye Yang ~
  Jiace Zhao ~
  Tianrui Liu ~ 
  Han Cao ~
  Yeyang Zhou \\ [0.3em] 
  Rui Wang ~
  Song Wang$^{5}$ ~
  Xiang Ren ~
  Bo Zhang ~
  Yutong Ban ~
  Pieter Abbeel$^{6}$ ~ 
  Brian Anthony$^{1}$ \\ [0.3em]
  $^{1}$MIT ~
  $^{2}$UIUC ~
  $^{3}$CMU ~ 
  $^{4}$USC ~ 
  $^{5}$UVA ~
  $^{6}$UC Berkeley \\ [0.3em]
  $^{\alpha}$ Equal Contribution ~ $^{\nu}$ Lead Researcher ~ 
  Correspondence: qinzy@alum.mit.edu
}

\maketitle

\begin{figure}[h]
    \centering
    \includegraphics[width=0.9\linewidth]{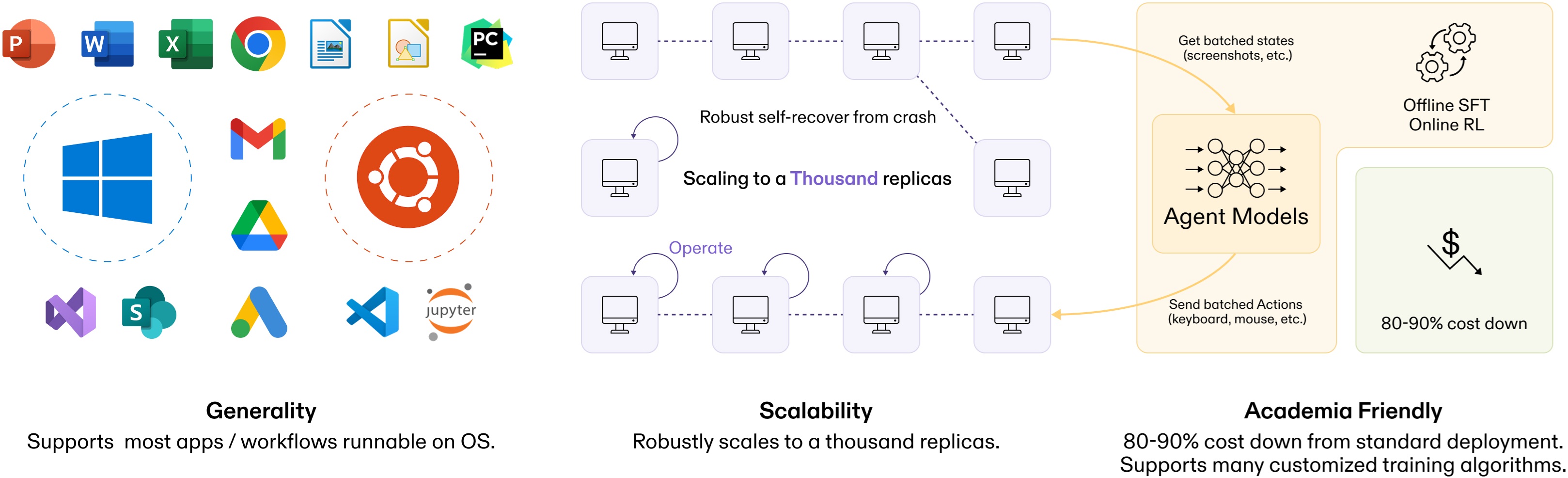}
\end{figure}

\begin{abstract}

Training computer use agents requires full-featured OS sandboxes with GUI environments, which consume substantial hardware resources as the number of sandboxes scales. Stochastic errors arising from diverse software execution within these sandboxes further demand robust infrastructure design and reliable error recovery. We present OSGym, a scalable OS environment infrastructure for computer use agents, built around these key optimization strategies: \textbf{(1) Decentralized OS state management}, which isolates failures to individual replicas and significantly enhances overall system reliability; \textbf{(2) Hardware-aware OS replica orchestration}, which addresses CPU-bounded scaling bottlenecks and substantially reduces compute overhead; \textbf{(3) KVM virtualization with copy-on-write disk management}, which shares a common bootable disk across VM instances and provisions only instance-specific modifications, reducing physical disk consumption by \textbf{88\%} and increasing disk provisioning speed by \textbf{37 times}; and \textbf{(4) Robust container pool with multi-layer fault recovery}. Together, these optimizations yield strong scalability and resource efficiency: OSGym manages \textbf{over a thousand OS replicas} under constrained resources, supports parallel trajectory generation at \textbf{1420} multi-turn trajectories per minute, and reduces per-replica cost to \textbf{0.2–0.3 USD per day, a 90\% reduction} over standard deployment. Our experiments validate OSGym across end-to-end pipelines for data collection and training for computer use agents. We believe OSGym establishes a new foundation for scalable, general-purpose computer use agent research.

\end{abstract}

\thispagestyle{firstpage}

\section{Introduction}

Computer Use~\cite{openaicomputeruse,qin2025uitars,agashe2024agents1,agashe2025agents2,anthropicclaude,xu2024aguvis,hong2024cogagent,cheng2024seeclick,wu2024osatlas,gou2025navigatingthedigital} is an emergent capability of agentic foundation models, where the model can take the computer screen as input, operate the keyboard and mouse, and complete complex tasks across diverse software and web environment. Such capability implies massive potential value and may define agentic operating system. 

However, training computer use agents is highly non-trivial, especially in terms of the infrastructure. The training requires massive amount of agent-environment interactions across diverse scenarios in full-featured OS sandboxes with GUI~\cite{xie2024osworld,wu2024oscopilot}, which is much heavier than vertical sandboxes like a coding environment~\cite{li2022alphacode,Chen2021Codex}, command-line terminals~\cite{yang2023intercode,qiao2023taskweaver}, or web browsers~\cite{chezelles2024browsergym,deng2023mind2web,zhou2023webarena}. Scaling up the amount of such OS sandboxes would consume substantial CPU, RAM and disk resources. Errors arising from diverse software and web execution within these sandboxes further demand robust infrastructure design and reliable error recovery mechanism. Scaling to thousands of instances without careful management and optimization leads to degraded performance and cascading failures. Hosting even a few hundred OS environments on cloud infrastructure is expensive for academia research, making cost a practical bottleneck if without careful resource optimization.

We introduce OSGym, a scalable OS environment infrastructure for training computer use agents, built around the following core design principles. \textbf{Decentralized state management:} we assign each OS replica its own state manager responsible for handling state transitions, monitoring health, and recovering from failures autonomously. This isolates faults at the replica level and prevents error propagation across the system. \textbf{Hardware-aware orchestration:} we recognize that under identical hardware resources, different orchestration strategies hit different bottlenecks. We find that binding the system to RAM rather than CPU as the primary constraint enables superior scaling. \textbf{KVM virtualization with copy-on-write disk management:} we partition bootable disks into blocks shared across sandboxes, provisioning only the modified blocks per sandbox. This reduces physical disk usage by 88\% and accelerates bootable disk provisioning by 37 times. \textbf{Robust container pool with multi-layer fault recovery:} we maintain a fixed-size pre-warmed runner pool to accelerate the sandbox creation, carefully tuned the kernel parameters to prevent silent failure, and implemented step-level retry along with task-level runner reassignment to further ensure that isolated replica failures do not propagate into training disruptions.

Beyond infrastructure, OSGym supports a broad and diverse range of OS tasks, unifying their execution into a unified stage-wise pipeline. A centralized data server with a single entry point bridges the training loop with the data generation loop, streamlining end-to-end workflows. These design choices yield substantial gains in scalability, resource efficiency, and robustness. OSGym can launch and manage over a thousand OS sandboxes under typical academic resource constraints, supporting tasks spanning web browsing, document editing, software engineering, and multi-app workflows. It collects approximately 1,420 multi-turn trajectories per minute, while keeping per-sandbox costs to 0.20–0.30 USD per day on standard on-demand compute, a 90\% reduction from standard deployment costs. The rest of this paper will describe the design principles behind each of these properties in detail.

\begin{figure}
    \centering
    \includegraphics[width=\linewidth]{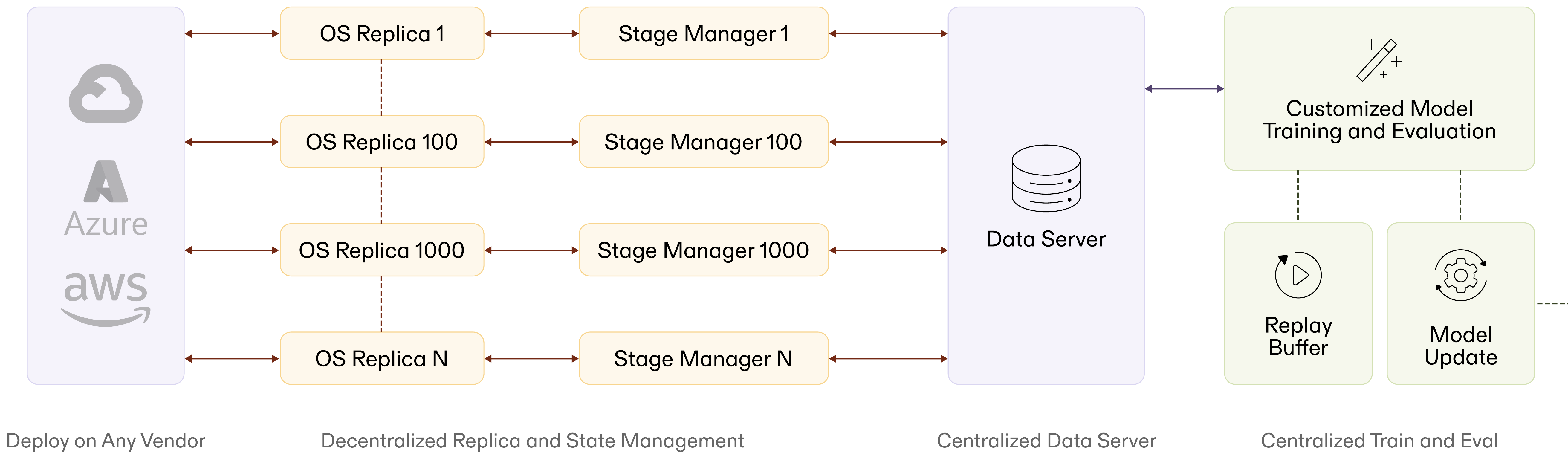}
    \caption{\textbf{OSGym Overview.} OSGym decentralizes the OS replica running and state management to achieve high scalability, without sacrificing the average performance of each replica when scaling to a thousand replicas. It also has robust fault tolerance mechanism so that failures in some replicas do not affect the whole. OSGym also supports a wide variety of tasks as long as they run on an operating system, which is important for training general-purpose computer agents. OSGym also has a centralized data server with single-entry interface exposed to the user, which hides the underlying complexity and is easy to use. OSGym is also algorithm-independent, compatible with customized training and evaluation loops. Lastly, OSGym can be deployed on any cloud providers and costs as low as 0.2 to 0.3 USD / replica / day (or free for self-hosting), making it affordable for academia use.}
    \label{fig:main}
\end{figure}

\section{Related Work}
\textbf{Web-Based Agents in Vertical Domains.} A large portion of the prior work on LLM agents has focused on vertical environments~\cite{wu2024oscopilot,Wang2024OSCAR,zhou2023webarena,Shi2017WoB,Wang2023Voyager,Zhang2024UFO,Zhang2024MMINA}, particularly web-based and coding-based tasks. For example, BrowserGym~\cite{chezelles2024browsergym}, WebArena~\cite{zhou2023webarena}, VisualWebArena~\cite{koh2024visualwebarena}, and WebVoyager~\cite{he2024webvoyager} offer environments where agents interact with websites using structured DOM interfaces or rendered browser views. While effective for benchmarking web navigation, these environments are inherently limited in scope because agents mainly work in a browser window and are not targeted at the broader operating system or performing multi-application workflows. Similar web-centric frameworks like WorkArena~\cite{drouin2024workarena}, WebRL~\cite{qi2024webrl}, and AgentLab~\cite{chezelles2024browsergym} further reflect these constraints.

\textbf{General-Purpose OS Environments.} To push beyond vertical agents~\cite{wu2024oscopilot,Wang2024OSCAR,zhou2023webarena,Shi2017WoB,Wang2023Voyager,Zhang2024UFO,Zhang2024MMINA,Tan2024rdr2,song2024beyond,Nakano2022WebGPT} and mobile-platform agents~\cite{Zhang2024LLAMATouch,Zhang2024Android}, OSWorld~\cite{xie2024osworld} and Windows Agent Arena~\cite{bonatti2024windowsagentarena} provide more realistic full OS environments. OSWorld introduces a diverse benchmark spanning office, browser, and developer tasks in real Linux environments, while Windows Agent Arena targets Windows-specific workflows. These efforts highlight the need for agents to operate in unrestricted digital environments, but neither system offer a scalable framework for high-throughput training or experimentation. They are primarily benchmark-oriented, lacking built-in support for rollout orchestration, resource scaling, or training integration.

\textbf{LLM-Based Generalist Agents.} Recent models such as OpenAI Operator~\cite{openaicomputeruse}, Claude Computer-Use~\cite{anthropicclaude}, Agent-S~\cite{agashe2024agents1}, Agent-S2~\cite{agashe2025agents2}, UI-TARS~\cite{qin2025uitars}, CogAgent~\cite{hong2024cogagent}, and Aguvis~\cite{xu2024aguvis} aim to train general-purpose agents capable of using software through language, vision, and API-calls~\cite{Shen2023HuggingGPT,Kim2023Solve,Humphreys2022Control}. These models explore instruction-following, thought-action decomposition, and GUI grounding across a range of benchmarks. Some, like OS-ATLAS~\cite{wu2024osatlas}, focus on building reusable action models, while others, like AutoGLM~\cite{liu2024autoglm}, emphasize multi-modal coordination. OSGym provides a systematic solution to the infrastructure problem, allowing future agents to be trained and evaluated in arbitrary software contexts, under realistic OS conditions, and at scalable throughput.

\section{OSGym: Scalable, Generalizable and Academia-Affordable}

\begin{figure}
    \centering
    \includegraphics[width=0.992\linewidth]{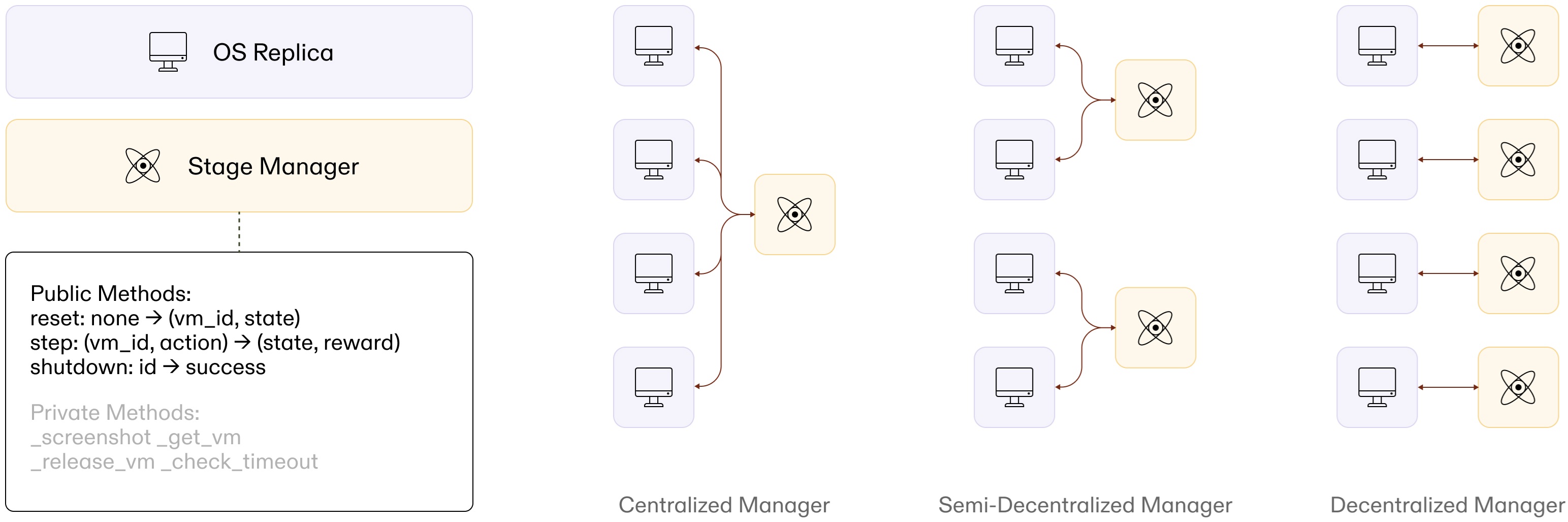}
    \caption{\textbf{Decentralized OS State Management.} In centralized state management, a single manager manages all OS replicas. In semi-decentralized state management, OS replicas are split into groups where each group is controlled by a single manager. In decentralized state management, each OS replica has its own stage manager. The state manager has public methods similar to OpenAI Gym~\cite{openaigym}, with a special set of private methods to low-level manage the state and healthiness of OS replicas.}
    \label{fig:decentralized_manager}
\end{figure}

\subsection{Decentralized OS State Management}

It is natural to consider three design options for the state manager: centralized, semi-decentralized, and decentralized, as illustrated in Figure~\ref{fig:decentralized_manager}. Full centralization introduces a critical performance bottleneck and poses significant risks to robustness. As the number of OS replicas scales into thousands, the centralized manager quickly becomes overwhelmed, leading to increased latency, reduced responsiveness, and an increased risk of single-point failures that can halt the entire system. In the semi-decentralized alternative, inter-group coordination still requires complex communication mechanisms, which may introduce delays and synchronization challenges, limiting scalability. OSGym adopts a fully decentralized design, where each OS replica has its own dedicated state manager. This architecture achieves optimal scalability and robustness, effectively eliminating bottlenecks associated with centralized control. Individual managers handle state transitions, monitor health, and recover autonomously from local failures. This isolation ensures that failures in one replica do not propagate, greatly enhancing system reliability and ease of maintenance.

\subsection{Hardware-Aware Optimization of OS Replica Orchestration}

\begin{figure}
    \centering
    \includegraphics[width=\linewidth]{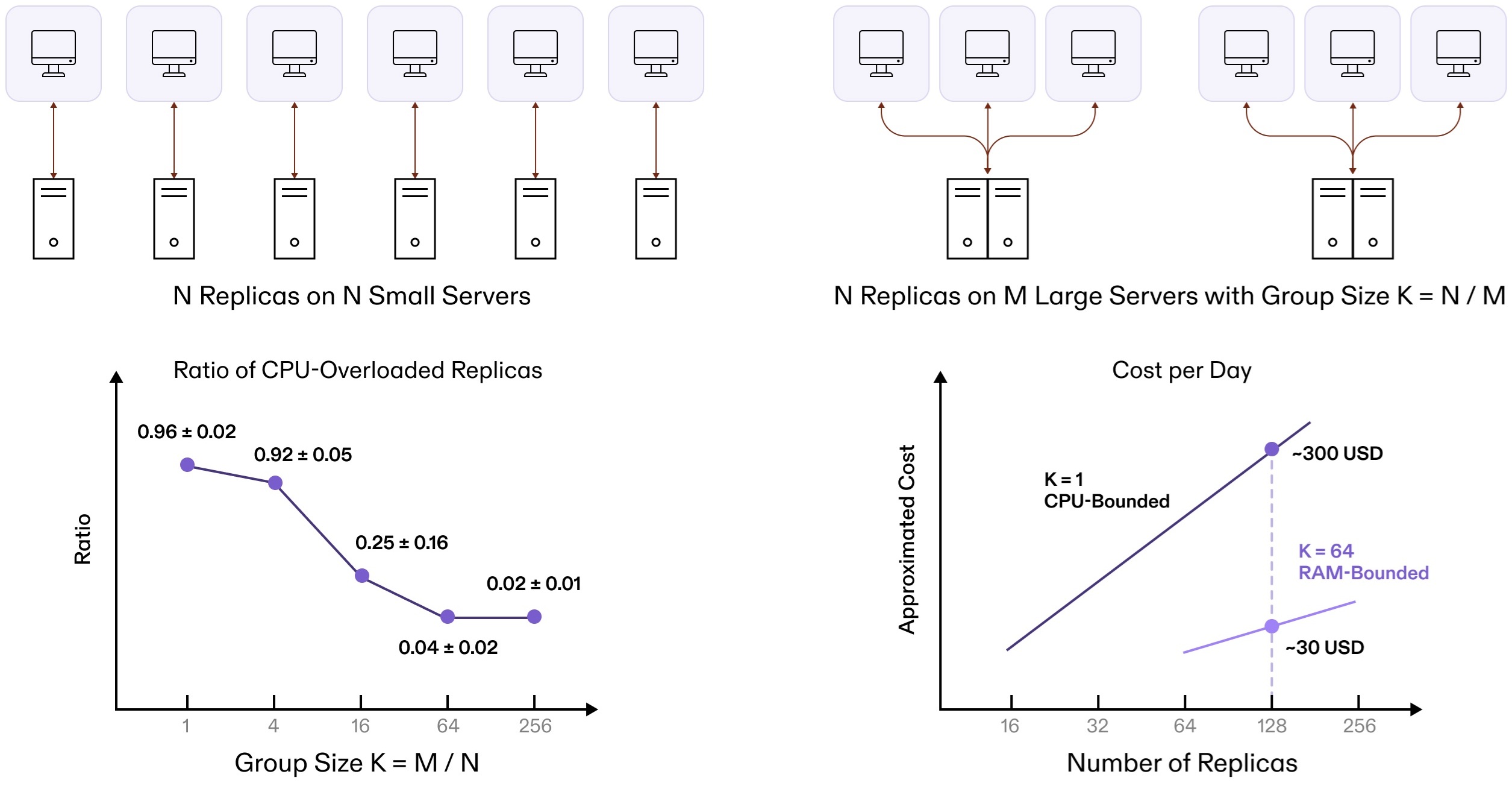}
    \caption{\textbf{Hardware-Aware Optimization of OS Replica Orchestration.} To cloud-deploy or self-host a large number of OS replicas, one may choose to host N replicas on N small servers, or on M large servers where each server hosts K = N / M replicas. We provide a useful insight that \textit{for small K, the scaling is CPU-bounded, while for large K, the scaling is RAM-bounded (see the bottom-left plot), and RAM is much cheaper than CPU.} So we increase the RAM of each server to use a large K, which significantly cuts down the cost (see the bottom-right plot). The numbers following ± represents the standard deviation across 10 independent runs.}
    \label{fig:semi_decentralized_orchestration}
\end{figure}

As we know, running full-fledged OS sandboxes with GUI environment can consume a non-trivial amount of computing resource. We choose to run replicas as Dockers rather than Virtual Machines to save the per-replica resource, and we found that the Docker images provided by OSWorld~\cite{xie2024osworld} are a good starting point. Cloud-hosting or self-hosting the OS replicas requires a large number of virtual or physical CPU servers, and the maximizing resource usage can significantly improve horizontal scaling. One option is to spread N replicas into N small servers, and the second option is to group the replicas and host each group on a larger server, as illustrated in Figure~\ref{fig:semi_decentralized_orchestration}. Let N be the number of OS replicas, and M be the number of servers, then K = N / M is the group size. We provide a useful insight regarding the different bottlenecks faced by the system under different K:

\begin{remarkbox}
For a small K, the OS replica scaling is usually bounded by CPU resource. For a large K, the scaling is usually bounded by memory (RAM) resource instead.
\end{remarkbox}

To help the readers understand, we freeze N and change K and plot the two graphs at the bottom of Figure~\ref{fig:semi_decentralized_orchestration}. The bottom-left plot shows that for small K, almost every replica is CPU-overloaded. When changing to a large K, even though the total amount of CPU resource is unchanged, the CPU overload diminishes because different replicas usually have peak CPU usage at not completely overlapping time. \textbf{Under a large K, the bottleneck is no longer the CPU but the RAM. Scaling RAM is significantly cheaper than scaling CPUs}. A 32GB RAM with DDR4 is usually \textit{only 10\% to 20\%} of the price of a 16-core CPU. We also provide a more concrete example of the cost in the bottom right plot of Figure~\ref{fig:semi_decentralized_orchestration}. Under the current market price of cloud computer on-demand rental, assuming the CPU is Intel Xeon series, and the RAM is DDR4 from Samsung, and we run 128 OS replicas. The daily cost is around 300 USD if K = 1, but only around 30 USD if K = 64. Each replica costs around 30 / 128 = 0.234 USD per day. Typically, 128 replicas can already support a decent academia-scale experiment on agent training, and the cost perfectly fits into academia budget in many cases. We hope that this discovery can help many academia labs unlock at least a part of the scaling potentials in general-purpose computer agent research.

\subsection{KVM Virtualization with Copy-on-Write Disk Management}
\label{sec:cow}

A practical challenge in deploying a large amount of OS replicas is disk provisioning. Each VM requires its own bootable disk image of approximately 24~GB. Naively duplicating the base image for each replica would require terabytes of storage and minutes of provisioning time per VM, making large-scale deployment infeasible.

OSGym addresses this through filesystem-level \textbf{reflink copy-on-write} (CoW). Each per-VM disk image is created instantaneously with \texttt{cp --reflink=always} and shares physical disk blocks with the base image. \textbf{Blocks are only physically allocated when a VM writes to them. Only blocks modified by each VM are allocated.} Combined with KVM hardware virtualization via \texttt{/dev/kvm}, each replica boots from its own CoW copy of the base image with \textbf{near-zero disk overhead and near-native CPU performance}. We will show in the experiment section that our method \textbf{reduces disk consumption by around 88\%}, which means we can obtain the full logical disk by using only 12\% physical disk.

We provision storage on NVMe drives striped in RAID~0 or LVM, formatted as XFS, the only production-grade filesystem supporting both reflink copy-on-write and high-concurrency I/O. Several XFS performance tunings are applied, including large preallocation (\texttt{allocsize=1G}) to reduce fragmentation from concurrent \texttt{qcow2} writes, disabled access-time updates (\texttt{noatime}), and an enlarged log buffer (\texttt{logbsize=256k}). To counteract gradual XFS performance degradation under sustained high-concurrency VM I/O, OSGym optionally deploys a systemd timer that reboots idle nodes every 48 hours, keeping the filesystem in peak condition.

\subsection{Robust Container Pool with Multi-Layer Fault Recovery}

Creating and destroying VM instances on-demand is prohibitively slow and fragile: a single failure in the creation-execution-teardown chain can stall an entire training batch. Also, 
running 128 VMs per node quickly exhausts default kernel resource, causing silent failures under high concurrency. 

OSGym addresses these challenges through the following design: \textbf{Pre-Warmed Runner Pool:} Each executor node maintains a fixed-size runner pool (default: 128 runners) initialized at startup. Rather than provisioning VMs on demand, OSGym pre-creates all instances before training begins and recycles them 
between tasks: upon task completion, each runner is reset and returned to the pool, ready for immediate reassignment. \textbf{Resource Guard:} Before each VM creation, OSGym reads \texttt{/proc/meminfo} and \texttt{/proc/loadavg} to verify that the host can safely accommodate another instance. Creation is blocked if available memory falls below 10\% or under 8~GB 
absolute. To prevent over-provisioning under burst scenarios, OSGym tracks in-flight creations and subtracts their estimated memory footprint (each container is limited to 6~GB) before evaluating available headroom. \textbf{System Limits Tuning:} Running 128 VMs within Docker containers requires explicit kernel parameter tuning. Default system limits, including file 
descriptors, inotify watches, AIO contexts, and netfilter connection tracking, are far too low for high-concurrency VM workloads and cause silent failures without adjustment. OSGym raises these limits to production-appropriate values (e.g., \texttt{fs.aio-max-nr} from 65,536 to 1,048,576; \texttt{fs.inotify.max\_user\_instances} from 128 to 8,192). \textbf{Multi-Layer Retry and Error Recovery:} OSGym implements fault tolerance at two levels. At the step level, each action execution is wrapped with a configurable retry policy (default: 10 retries) covering connection errors, timeouts, and runtime 
operation failures. At the task level, if a runner fails permanently, the task is reassigned to a fresh runner from the pool; leaked tasks that exceed a timeout threshold are automatically reclaimed and their runners returned to circulation. A gateway layer further provides task-affinity routing, background health checks every 10 seconds, and automatic failover when an executor node becomes unreachable. 
Together, these mechanisms ensure that isolated replica failures do not propagate into training disruptions.

\subsection{Universally Diverse Tasks with Unified Flow}

\begin{figure}[ht]
    \centering
    \includegraphics[width=\linewidth]{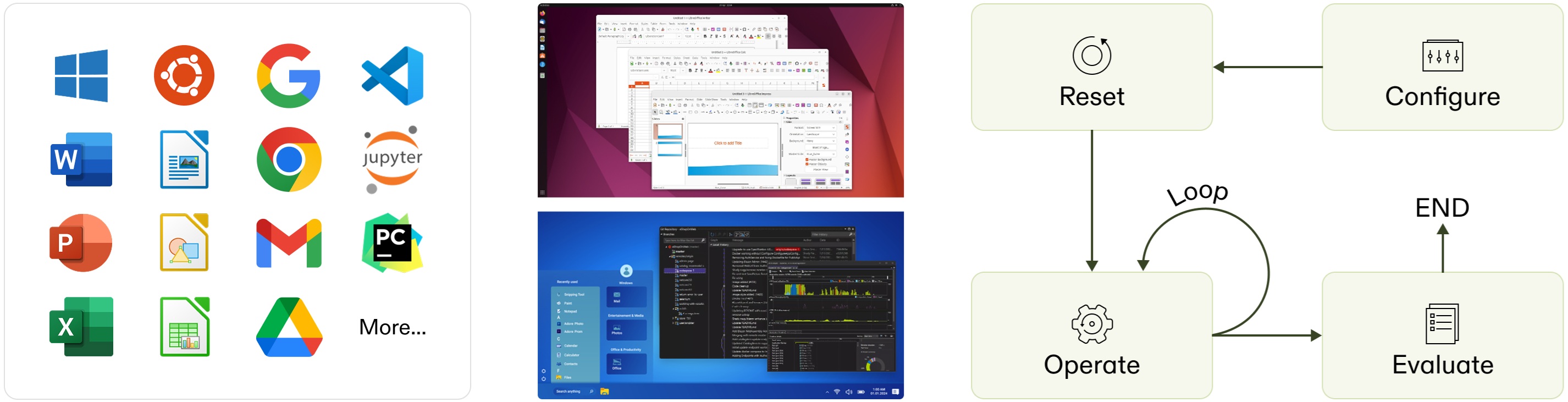}
    \caption{\textbf{Diverse Tasks with Unified Flow.} Since OSGym does not run specialized sandbox but runs full-fledged OS, it naturally supports a wide variety of tasks as long as the involved software run on the OS. OSGym also unifies the operation flow where each task has 4 parts, configure, reset, operate and evaluate, controlled by the public methods of the state manager.}
    \label{fig:diverse_tasks}
\end{figure}

OSGym inherently supports an extensive range of tasks thanks to its deployment of fully operational OS replicas, rather than specialized, constrained sandboxes. Tasks from diverse software domains, such as software engineering (e.g., code debugging, software testing), office applications (e.g., word processing, spreadsheet manipulation), internet browsing, tool-based interactions, file management, and even complex multi-software workflows, can all be naturally supported within the unified OSGym infrastructure. OSGym adopts a unified execution flow comprising four consistent phases: \textbf{1) Configure.} Setting up necessary software, and preparing the OS environment with customized conditions. \textbf{2) Reset.} Before executing a task, the OS environment is reset to the initial conditions defined during the configuration, ensuring reproducibility and consistency between runs. \textbf{3) Operate.} The agent interacts with the OS through actions such as keyboard inputs, mouse movements, clicks, and potentially API-driven tool interactions, driven by observations typically captured through screenshots or additional metadata extracted from the OS. \textbf{4) Evaluate.} OSGym evaluates outcomes based on predefined criteria or metrics. We give the user full flexibility to customize the evaluation function, and call the evaluation function whenever necessary. We illustrate this section in Figure~\ref{fig:diverse_tasks}.

\subsection{Centralized Data Server with Easy-to-Use Single Entry}

\begin{figure}
    \centering
    \includegraphics[width=\linewidth]{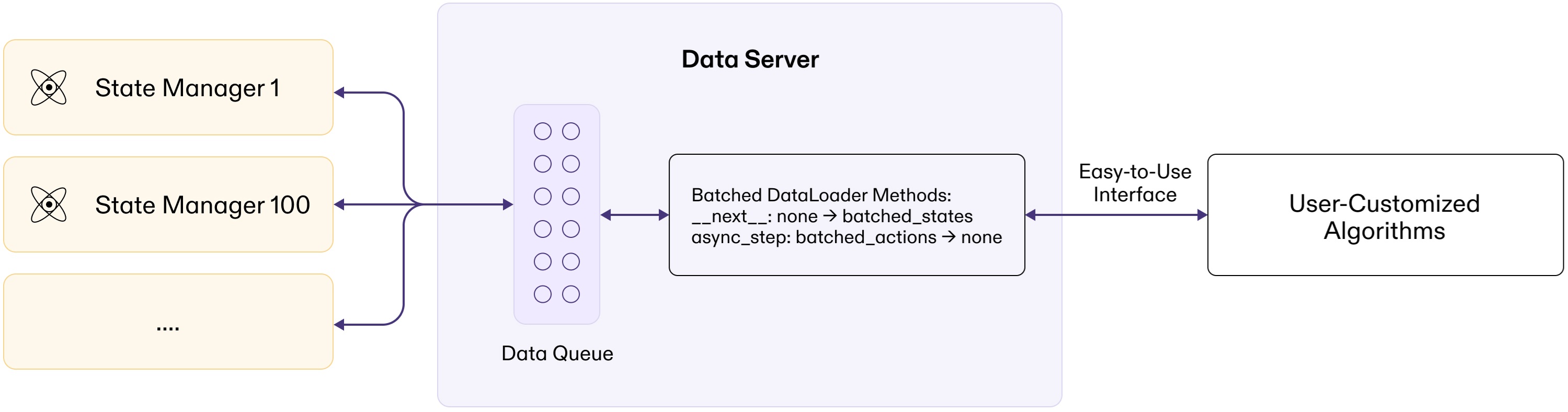}
    \caption{\textbf{Centralized Data Server with Easy-to-Use Single Entry.} The data server is easy-to-use with single-entry batched methods. The complexities of state manager communication and data queuing is internally managed by the data server. The batched step method in the data server is designed to be asynchronous so that the training or evaluation loop is not blocked.}
    \label{fig:centralized_dataloader}
\end{figure}

OSGym has a high-level centralized data server Python Class that provides an intuitive, single-entry interface to simplify interactions and data handling across numerous parallel OS replicas. The centralized data server manages all internal communications and queuing complexities with state managers, thus abstracting the low-level details away from the end-users. We illustrate the data server in Figure~\ref{fig:centralized_dataloader}. The key features of this centralized data server include: \textbf{1) Single Entry Interface}: Offers straightforward, batched methods such as reset and step, making the interaction with multiple OS replicas seamless and easy. \textbf{2) Asynchronous Operations}: The step method supports asynchronous execution, preventing blocking behavior during training or evaluation loops, significantly enhancing overall efficiency. \textbf{3) Internal Queuing and Management}: Automatically handles task queuing, replica availability checks, and dynamic load balancing, thereby maintaining system stability and scalability. \textbf{4) Fault Tolerance and Recovery}: Includes built-in error-handling capabilities to quickly recover from replica failures without interrupting overall service availability.

\section{Experiments}

In the experiments section, we mainly target at evaluating the scalability, robustness and resource efficiency of OSGym. We will also implement an example training pipeline to demonstrate its practical usefulness.

\begin{figure}[ht]
    \centering
    \includegraphics[width=\linewidth]{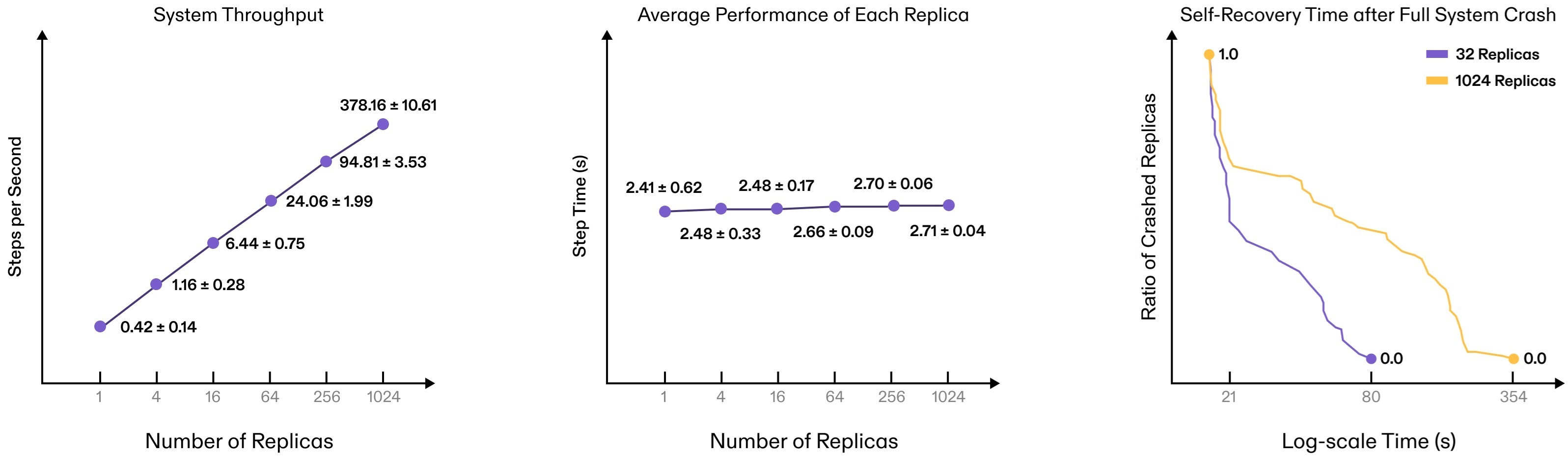}
    \caption{\textbf{Scalability and Robustness Analysis.} The left figure shows a near-perfect linear scaling of system throughput with increasing number of replicas. The middle figure shows that the average performance of each replica is maintained even though we significantly scale up the system size. The right figure shows results of a robustness test, where the system starts from full crash and manages to completely self-recover within acceptable time. The numbers following ± represent the standard deviation across 10 independent runs.}
    \label{fig:scalability_analysis}
\end{figure}

\subsection{Scalability, Robustness and Resource Efficiency Analysis}

\begin{table}[ht]
  \caption{\textbf{CPU Machine Specifications and OSGym Hosting.} The table shows three types of cloud CPU machines and compares the hosting cost of OSGym. A large-RAM machine generally costs less than a high-end-CPU machine. When using Intel E5-2699 CPU (88 cores) and 768~GB DDR4 RAM, the average hosting cost per replica per day is as low as 0.23 USD.}
  \label{tab:replica_costs}
  \centering
  \begin{tabular}{cccccc}
    \toprule
    CPU Cores & RAM & CPU Type & RAM Type & Replicas per Machine & Cost per Replica \\
    \midrule
    96 & 192 GB & 8275CL & DDR4 & 36 & 2.10 USD / day \\
    96 & 768 GB & 8259CL & DDR4 & 128 & 0.78 USD / day \\
    88 & 768 GB & E5-2699 & DDR4 & 128 & \textbf{0.23 USD} / day \\
    \bottomrule
  \end{tabular}
\end{table}

\textbf{Scalability.} A crucial metric to examine the scalability of a system is whether its throughput proportionally increases with the parallelization size. It is not uncommon to see diminishing returns in large-scale systems, where increasing the system size fails to yield proportional gains in throughput due to bottlenecks, resource contention, or system overhead. OSGym, in contrast, demonstrates highly favorable scalability. As shown in the left plot of Figure~\ref{fig:scalability_analysis}, the system throughput, measured in steps per second, increases nearly linearly with the number of OS replicas. This indicates that OSGym scales efficiently across a wide range of deployment sizes, from tens to thousands of environments.

Further, the middle plot of Figure~\ref{fig:scalability_analysis} reveals that the average step latency per replica experiences only a marginal increase as the number of concurrent replicas grows exponentially. This is a strong indication that OSGym's decentralized management architecture and semi-decentralized orchestration strategy successfully mitigate common scaling pitfalls. It ensures that each environment continues to operate with minimal degradation even under heavy load. Taken together, these results provide solid evidence of OSGym's strong scalability, making it suitable for both small-scale experimental setups and large-scale training infrastructures. The system maintains high throughput and reliability across different scales, a critical requirement for sustained and efficient training of general-purpose agents in complex operating system environments.

\textbf{Robustness.} In large-scale distributed systems, robustness is critical to ensuring sustained functionality in the presence of inevitable faults. OS replicas can encounter a wide range of stochastic failures due to software bugs, kernel crashes, system misconfigurations, or network issues. If left unhandled, such failures can accumulate and ultimately halt the entire system. To address this, OSGym integrates a decentralized self-recovery mechanism within each OS state manager. When a replica encounters a critical error or becomes unresponsive, its local manager detects the failure, isolates the faulty instance, and autonomously initiates a recovery procedure. As shown in the right plot of Figure~\ref{fig:scalability_analysis}, even when the system is initialized in a fully crashed state, OSGym is capable of self-restoring all replicas to a healthy condition within a short recovery window. This high degree of robustness is essential for maintaining long-term, uninterrupted agent training and evaluation at scale.

\textbf{Resource Efficiency.} As shown in Table~\ref{tab:replica_costs}, careful selection of server configurations, particularly those with high memory capacity, allows substantial cost savings when running OSGym at scale. By hosting multiple OS replicas on large-RAM servers, we significantly reduce the per-replica cost. For example, using a server with an 88-core Intel E5-2699 CPU and 768~GB DDR4 RAM, the cost per OS replica can be brought down to just 0.23 USD per day. This makes large-scale experimentation with hundreds of replicas financially feasible for academic labs. This cost efficiency makes it practical for both academic and commercial users to pursue research and development of general-purpose computer agents without excessive infrastructure expenses.

We also evaluate the effectiveness of our KVM virtualization with copy-on-write disk management. As explained in Section~\ref{sec:cow}, since each OS replica VM requires its own bootable disk, launching a large amount of VM instances on a bare metal machine would consume huge physical disk if there is no optimization. With reflink copy-on-write, we share the unmodified common blocks across the VMs and only provision the modified blocks for each VM instance. The result is a 88\% reduction of physical disk usage and 37 times faster disk provisioning speed as shown in Table~\ref{tab:reflink}. The logical disk of each instance is not affected, and it also gives near-native CPU performance.

\begin{table}[ht]
\caption{\textbf{Disk Consumption With and Without Reflink Copy-on-Write.} Reflink sharing reduces physical disk consumption by 88\% across 128 VM instances on a bare metal machine.}
\label{tab:reflink}
  \centering
  \begin{tabular}{lcc}
    \toprule
    \textbf{Metric} & \textbf{Without Reflink} & \textbf{With Reflink} \\
    \midrule
    Per-VM disk provision time       & 30s   & \textbf{0.8s (37 times faster)} \\
    Physical disk for 128 VMs     & 3.1 TB & \textbf{366 GB (88\% reduction)} \\
    Logical disk per VM  & 24 GB & 24 GB (no degradation) \\
    \bottomrule
  \end{tabular}\end{table}

\begin{table}[ht]
  \caption{\textbf{Data Generation Statistics with Example Pipeline Implemented with OSGym.} This table summarizes the number of valid trajectories and steps collected across different application domains. The tasks span Office, Daily, and Professional use cases, as well as combined workflows. Effective generation times are also provided, showing a significant speedup with parallelization.}
  \label{tab:osgym_tasks}
  \centering
  \begin{tabular}{cllcc}
    \toprule
    \textbf{\quad Task Type \quad} & \textbf{Domain} & \textbf{Description} & \textbf{Trajectories} & \textbf{\quad Steps \quad} \\
    \midrule
    \multirow{3}{*}{Office} 
      & LibreOffice Writer & Document Editing       & 493 & 5028 \\
      & LibreOffice Calc   & Spreadsheet Editing    & 222 & 4240 \\
      & LibreOffice Impress& Presentation Editing   & 314 & 4898 \\
    \midrule
    \multirow{3}{*}{Daily} 
      & Chrome             & Web Browsing           & 291 & 4285 \\
      & ThunderBird        & Email                  & 189 & 3627 \\
      & VLC                & Media Control          & 107 & 1701 \\
    \midrule
    \multirow{3}{*}{Professional} 
      & VS Code            & Programming            & 309 & 4604 \\
      & GIMP               & Image Editing          & 203 & 3410 \\
      & OS                 & System Configuration   & 491 & 5333 \\
    \midrule
    Workflow 
      & Multi-Apps         & Combined Above         & 244 & 5709 \\
    \bottomrule
  \end{tabular}

  \vspace{1em}
  \begin{tabular}{p{0.94\linewidth}}
    \toprule
    \textbf{Net Generation Time} (total time minus overhead such as machine setup): \\
    Without OSGym Parallelization: 115,654 seconds \\
    With OSGym 1024-Replica Parallelization: \textbf{121 seconds ($\approx$1420 trajectories / min)} \\
    \textbf{Net Cost on Cloud Machine Rental}: \textbf{43 USD} \\
    \bottomrule
  \end{tabular}
\end{table}

\subsection{Example Application of OSGym}\label{sec:example_training}

We used OSGym to implement an example pipeline to train computer-use agents. The pipeline includes highly-parallel data generation, supervised finetuning, and reinforcement learning.

\textbf{Data Generation with OSGym.}
We first manually prepared 244 task prompts following the style of OSWorld (but not overlapping with OSWorld original tasks), involving multiple software such as LibreOffice Writer / Calc / Impress, Chrome, GIMP, VLC, VS Code, and ThunderBird, spanning office tasks, professional tasks, daily tasks, multi-app workflow tasks, etc. Then we used existing open-source computer-use agents~\cite{agashe2025agents2, qin2025uitars} to run on these tasks to generate a large number of demonstration trajectories. Leveraging OSGym’s massive parallelization, we deployed 1024 OS replicas to execute and collect these demonstrations simultaneously, at an average speed of 1420 trajectories per minute. Each trajectory contains 10 to 25 steps of interleaved states, actions, and thoughts (reasoning) before each action. Thanks to the cost-efficient infrastructure provided by OSGym, the entire dataset was generated within minutes and at a total cloud cost of only 43 USD, making it highly accessible for academic-scale research.

\textbf{Supervised Finetuning.} After data generation, we finetuned the Qwen 2.5-VL 32B~\cite{bai2025qwen2} model on the collected data. Each data sample is structured as a sequence: task instruction $\rightarrow$ screenshot$_1$ $\rightarrow$ thoughts$_1$ $\rightarrow$ action$_1$ $\rightarrow$ screenshot$_2$ $\rightarrow$ thoughts$_2$ $\rightarrow$ action$_2$ $\rightarrow \dots \rightarrow$ screenshot$_C$ $\rightarrow$ thoughts$_C$ $\rightarrow$ action$_C$. For training, we conditioned the model on the initial task instruction and the history of prior elements (screenshots, thoughts, and actions). We then applied a softmax cross-entropy loss to the model prediction for each subsequent thought and action in the sequence. We trained the model using the Adam~\cite{kingma2014adam} optimizer with a learning rate of $10^{-5}$ until it converged, which took approximately half a day on a 8×H100 machine.

\textbf{Agent Reinforcement Learning with OSGym.} We performed reinforcement learning on the finetuned model using a semi-online asynchronous pipeline implemented with OSGym, where data rollouts and model updates are decoupled and run in parallel. This design maximizes resource utilization and training throughput by keeping the OS replicas continuously busy with interactions while the model updates run independently. For each interaction step, actions are predicted by the current model and dispatched to the corresponding OS replicas via OSGym's batched, asynchronous interface. The resulting experiences are added to a replay buffer, from which the model samples batches for policy and value updates using standard PPO objectives. The model was trained for 200 steps with batch size 64 and learning rate $10^{-6}$ using Adam~\cite{kingma2014adam} optimizer.

To evaluate the model trained with OSGym, we ran it on OSWorld-Verified benchmark with each task given a 100-step limit. The model achieves a success rate of 56.3, which is competitive with existing methods given that it uses a 32B parameter base model with no task-specific tuning. The goal of this experiment is not to establish a benchmark result, but to validate that OSGym supports an effective end-to-end training pipeline, from data collection through supervised finetuning to reinforcement learning, and that the resulting model is a functional computer-use agent.

\section{Limitations, Discussions and Broader Impacts} \label{sec:discussion}

OSGym presents a practical and scalable infrastructure, but we believe there are several limitations that are important to acknowledge to clarify the scope and to motivate future work. The first is on Task Collection and Reward Modeling. Although OSGym supports general tasks that run on an OS, creating high-quality tasks with reliable reward functions remains nontrivial. Many OS tasks involve multiple applications, file manipulations, or UI subtleties that are hard to formalize into success criteria. While the system provides a flexible interface for defining reward functions, researchers still need to invest time in curating tasks and crafting evaluation logic for new domains. Developing a library of standardized, community-contributed tasks and metrics would help resolve this limitation. The second is on Lack of Real-Time Human Feedback. Human-in-the-loop training remains underexplored within the current framework. Integrating real-time human feedback, such as through preference modeling or interactive corrections, may significantly improve agent performance and robustness, especially on open-ended tasks with ambiguous goals.

We hope OSGym can contribute to the development of society-wide accessible general-purpose computer agents that boost societal productivity. But we are also aware that such agent models can also be exploited for unintended use such as cyberattacks. Therefore, it is crucial to approach their development and deployment with a strong emphasis on safety, transparency, and ethical considerations.

\newpage

\bibliographystyle{acm-sigchi}
\bibliography{neurips_2025}
\end{document}